\documentclass[pra,twocolumn,superscriptaddress,amsmath,amssymb]{revtex4}
\usepackage{graphicx}% Include figure files
\usepackage{bm}
\usepackage{amsmath,amssymb}
\usepackage{color}
\usepackage{ascmac}

\begin{document}
\title{Composable security of measuring-Alice blind quantum
computation}  
\author{Tomoyuki Morimae}
\email{morimae@gunma-u.ac.jp}
\affiliation{ASRLD Unit, Gunma University,
1-5-1 Tenjin-cho Kiryu-shi Gunma-ken, 376-0052, Japan}

\author{Takeshi Koshiba}
\email{koshiba@mail.saitama-u.ac.jp}
\affiliation{Graduate School of Science and Engineering,
Saitama University, 255 Shimo-Okubo, Sakura, Saitama 338-8570, Japan}

%\author{Keisuke Fujii}
%\email{keisukejayorz@gmail.com}
%\affiliation{The Hakubi Center for Advanced Research, Kyoto University,
%Yoshida-Ushinomiya-cho, Sakyo-ku, Kyoto 606-8302, Japan}
%\affiliation{Graduate School of Informatics, Kyoto University,
%Yoshida Honmachi, Sakyo-ku, Kyoto 606-8501, Japan}
%\affiliation{Graduate School of Engineering Science, Osaka University,
%Toyonaka, Osaka 560-8531, Japan}
\date{\today}
            
\begin{abstract}
Blind quantum computing 
[A. Broadbent, J. Fitzsimons, and E. Kashefi, 
Proceedings of the 50th Annual IEEE Symposium on Foundations of Computer 
Science 517 (2009)]
is a secure cloud quantum computing
protocol which enables a client 
(who does not have enough quantum technology at her disposal)
to delegate her quantum computation to a server (who has a universal
quantum computer) without leaking any relevant information to the server. 
In [T. Morimae and K. Fujii, 
Phys. Rev. A {\bf87}, 050301(R) (2013)],
a new blind quantum computing protocol, so called 
the measuring-Alice protocol,
was proposed. This protocol
offers several advantages over previous protocols,
such as the device-independent security, 
less demanding requirements for the client, and
a simpler and stronger security based on the no-signaling principle.
In this paper, we show composable security
of the measuring-Alice protocol by using the formalism
of the constructive cryptography [U. Maurer, 
Proceedings of
Theory of Security and Applications, TOSCA 2011, pages 33-56. Springer
(2011)
].
The above advantages of measuring-Alice protocol
enable more intuitive and transparent proofs for
the composable security.
\end{abstract}

\maketitle  
%------------------------------------
\section{Introduction}
A first generation quantum computer will be expensive and high-maintenance,
and therefore will be
implemented in a ``cloud" style like today's supercomputers.
In such a cloud quantum computing,
the most important problem
is to guarantee the client's privacy. 
Blind quantum computation~\cite{BFK,FK,Barz,Vedran,AKLTblind,topoblind,CVblind,topoveri,MABQC,Sueki,composable}
is a new secure quantum computing protocol which
can protect the security of client's privacy
in such a cloud quantum computing.
Protocols of blind quantum computation
enable a client (Alice), who does not have enough quantum technologies
at her disposal,
to delegate her quantum computation to a server (Bob), who has
a full-fledged quantum computer,
in such a way that Alice's input, output, and program are hidden
to Bob~\cite{BFK,FK,Barz,Vedran,AKLTblind,topoblind,CVblind,topoveri,MABQC,Sueki,composable}.

An unconditionally secure protocol of blind quantum computation 
with almost classical Alice was first proposed
by Broadbent, Fitzsimons, and Kashefi (BFK)~\cite{BFK}.
Their protocol uses measurement-based quantum
computation (MBQC) on the cluster state (graph state) 
by Raussendorf and Briegel~\cite{Raussendorf}.
A proof-of-principle experiment of the BFK protocol
has also been achieved recently with a quantum optical system~\cite{Barz}.
The BFK protocol has been generalized to
other blind quantum computing
protocols which use MBQC
on the Affleck-Kennedy-Lieb-Tasaki (AKLT) state~\cite{AKLT,AKLTblind,Miyake},
continuous-variable MBQC~\cite{CV,CVblind},
topological MBQC~\cite{topoblind,Raussendorf_topo},
and the ancilla-driven model~\cite{Janet,Sueki}.

In these BFK-based protocols, Alice has to possess a device
which emits randomly-rotated 
single-particle states, such as single-photon states.
If Alice's device is not perfect, and sometimes wrongly
emits more than two photons,
extra photons can be exploited by malicious Bob to learn Alice's information,
by using, for example, the photon number splitting attack~\cite{PNS1,PNS2,
PNS3,PNS4}. 
Therefore, it is necessary for Alice to precisely control the
number of emitted photons,
which is not easy with today's technology.

In Ref.~\cite{MABQC}, 
a complementary protocol of the BFK-type protocol,
so called measuring-Alice (MA) protocol,
was proposed by Morimae and Fujii.
In this protocol,
Alice has only to perform measurements in stead of state preparations.
These measurements are not necessary to be single-photon
measurements: they can be polarization measurements with a threshold
detector.
Since a polarization measurement with a threshold detector
is much easier than a creation of a single-photon state,
MA protocol eases Alice's burden.
Furthermore, as is shown in Ref.~\cite{MABQC}, 
this protocol offers the device-independent security
for Alice, which means that Alice does not need to trust
her device: even if Alice's device does not work correctly,
Bob cannot learn Alice's information.
Finally, it was shown~\cite{MABQC} 
that the security of MA protocol 
is based on the no-signaling principle,
which is more fundamental than quantum physics~\cite{Popescu}.
Because of the no-signaling principle,
the proof of the security of MA protocol becomes very simple.

If Bob cannot learn anything about Alice's computation
whatever he does,
we say that the protocol offers ``blindness"~\cite{BFK,FK,Barz,Vedran,AKLTblind,topoblind,CVblind,topoveri,MABQC,Sueki,composable}.
(Here, we ignore unavoidable leakage of
trivial information, such as the upper bound of the computational
size, or whether Alice's output is classical or quantum, etc.~\cite{BFK,FK,Barz,Vedran,AKLTblind,topoblind,CVblind,topoveri,MABQC,Sueki,composable}.)
In fact, all BFK-type blind protocols and MA protocol
satisfy the blindness.

The ``verifiability" is another important concept in could quantum computing.
The verifiability means that Alice can check whether Bob is following
the correct protocol. If malicious Bob deviates from the correct protocol, 
the verifiability enables Alice to detect it with high probability,
and therefore the probability for Alice of accepting a wrong result
can be exponentially small.
Fitzsimons and Kashefi (FK) recently introduced a modified version
of the BFK protocol
which satisfies the verifiability~\cite{FK}.
The verifiability of 
MA protocol was shown in Ref.~\cite{topoveri}.

However, all previous results (except for Ref.~\cite{composable})
consider only stand-alone security.
The stand-alone security means that the protocol
is secure during a single execution of it in an isolated environment.
The stand-alone security is often proven by 
showing that the mutual information
between honest party and malicious party is exponentially
small.

The stand-alone security establishes the security
of a protocol as a primitive, and often gives important insights for
deep understanding of protocols.
Therefore, it is a first goal to show the stand-alone security
of a cryptographic protocol.
However, the stand-alone security is
not sufficient if we consider a protocol in a broader and hence more realistic
scenario.
For example, the (stand-alone)
unconditional security of the quantum key distribution
protocol (QKD)~\cite{QKD1,QKD2} has been proven by many researchers
by showing that the accessible information is exponentially small.
Here, the accessible information is 
the mutual information between the distributed key
and the outcome of an optimal measurement on the adversary's system.
However, it was pointed out in Ref.~\cite{Konig} that
even if the accessible information is small, the key might not be
enough secure if it is used in another protocol, such
as the one-time pad encryption,
due to the locking~\cite{locking}, which is a purely non-classical property.

If we want to guarantee the security of a protocol in 
such a broader context, we have to show the
composable security~\cite{MR11,Mau11,Unr09,Unr04,Canetti,BPW,UCQKD,BM02,BM04}.
The composable security means the security of a primitive protocol
in a general environment. 
For example, a protocol is secure
even if it is used many times as subroutines of a larger protocol.
The composable security of QKD
was shown in Ref.~\cite{UCQKD}.
The composable security of the key recycling in authentication
was studied in Ref.~\cite{Portmann}.

The composable security of the BFK-type protocols
were studied in Ref.~\cite{composable}. 
They showed the composable blindness of the BFK protocol
and the composable blind-verifiability of the FK protocol
by using the constructive cryptography~\cite{MR11,Mau11}.
Although the composable security of MA protocol
was also studied in Ref.~\cite{composable},
it is not sufficient, since
they showed only the composable blindness:
the composable device-independent blindness,
which is a new feature of MA protocol, was not shown.
Furthermore, the composable blind-verifiability of MA protocol
was neither considered.

In this paper, we study the composable security of MA protocol.
For that purpose, we 
utilize the framework of the constructive
cryptography~\cite{MR11,Mau11}.
We will introduce two types of MA protocols,
one is without verification and the other is with verification,
and will show
the composable security of them.
Hence, the device-independent blindness and the verifiability
of MA protocol are shown to be composable.
We will see that our proofs of 
the composable security
are much simpler than those of BFK type protocols
due to the simplicity of the stand-alone security
of MA protocol based on the no-signaling principle.

This paper is organized as follows. We will first review
some necessary backgrounds,
including the no-signaling principle (Sec.~\ref{sec:NS}),
MBQC (Sec.~\ref{sec:MBQC}),
MA protocol (Sec.~\ref{sec:MA}),
and
the constructive cryptography and the composable security (Sec.~\ref{sec:comp}).
We will then show our results
in Sec.~\ref{sec:result1}
and Sec.~\ref{sec:result2}.
The discussion is given in Sec.~\ref{sec:discussion}.

%%%%%%%%%%%%%%%%%%%%%%%%%%%%%%%%%%%%%%%%%%
\section{No-signaling principle}
\label{sec:NS}
\begin{figure}[htbp]
\begin{center}
\includegraphics[width=0.45\textwidth]{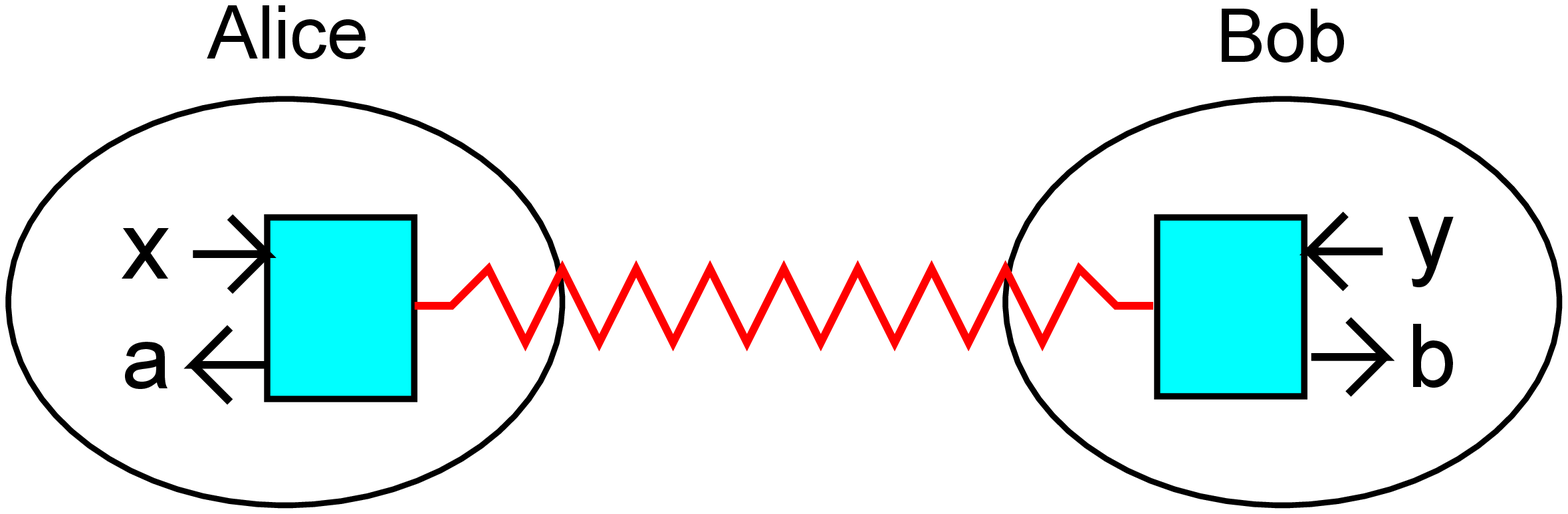}
\end{center}
\caption{
The no-signaling principle. Alice and Bob share a system.
} 
\label{NS}
\end{figure}

No-signaling principle is one of the most fundamental
principles in physics, and quantum theory also
respects it.
Formally, it is explained as follows.
Let us assume that Alice and Bob share a physical system, which might be
classical, quantum, or even super-quantum (Fig.~\ref{NS}).
For example, Alice and Bob share
the Bell pair,
\begin{eqnarray*}
\frac{1}{\sqrt{2}}(
|0\rangle_A\otimes|0\rangle_B+|1\rangle_A\otimes|1\rangle_B),
\end{eqnarray*}
where the subscript A (B) indicates Alice (Bob) possesses the qubit.

As is shown in Fig.~\ref{NS},
Alice chooses her measurement parameter $x$ (such as the measurement angles
of a spin), and performs measurement. She obtains the result $a$.
Bob also chooses his measurement parameter $y$, and performs
measurement. He obtains the result $b$.
The no-signaling principle (from Alice to Bob) is defined by
\begin{eqnarray}
P(b|x,y)=P(b|x',y)
\label{no-signaling}
\end{eqnarray}
for all $b$, $x$, $x'$, and $y$,
where $P(\alpha|\beta)$ is the conditional probability
distribution of $\alpha$ given $\beta$.
Equation~(\ref{no-signaling}) 
means that the change of Alice's measurement parameter
does not affect the probability distribution of Bob's measurement result.
In other words, the shared system cannot transmit
any message from Alice to Bob.

Interestingly, the no-signaling principle is more fundamental
than quantum theory in the sense that there is a theory
which is more non-local than quantum theory, but respects
the no-signaling principle~\cite{Popescu}.

%%%%%%%%%%%%%%%%%%%%%%%%%%%%
\section{Measurement-based quantum computing}
\label{sec:MBQC}
In this section, we will review the basics of measurement-based
quantum computing (MBQC)~\cite{Raussendorf}. Readers who are familiar with it
can skip this section.

MBQC is a new model of quantum
computation whose computational power is equivalent to
the traditional circuit model.
In MBQC, we first prepare a highly-entangled 
$N$ qubit (or more generally, qudit) state,
which we call the resource state.
We next perform measurement of each qubit. The measurement angle
of a qubit depends on the results of the
previous measurements.
If the resource state is a universal resource state,
we can simulate any quantum circuit with the adaptive
local measurements.

A canonical example of universal resource states
is the cluster state~\cite{Raussendorf}:
\begin{eqnarray*}
(\bigotimes_{(i,j)} CZ_{i,j})|+\rangle^{\otimes N},
\end{eqnarray*}
where $|+\rangle\equiv\frac{1}{\sqrt{2}}(|0\rangle+|1\rangle)$,
$N$ qubits are allocated on sites of the two-dimensional
square lattice, 
\begin{eqnarray*}
CZ_{i,j}\equiv |0\rangle\langle0|_i\otimes I_j+|1\rangle\langle1|_i\otimes Z_j
\end{eqnarray*}
is the controlled-$Z$ gate between $i$th qubit and $j$th qubit,
and $(i,j)$ is a pair of nearest-neighbour sites.
Here, $I_j$ is the identity operator acting on $j$th qubit,
and $Z_j$ is the Pauli $Z$ operator acting on $j$th qubit.

\begin{figure}[htbp]
\begin{center}
\includegraphics[width=0.48\textwidth]{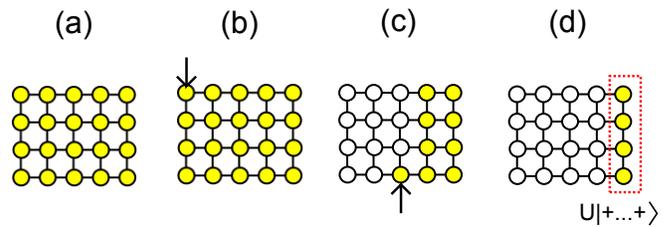}
\end{center}
\caption{
The measurement-based quantum computing with the cluster state.
} 
\label{figMBQC}
\end{figure}

Let us see Fig.~\ref{figMBQC}. We first prepare the cluster state
(Fig.~\ref{figMBQC} (a)).
We then perform local adaptive measurements (Figs.~\ref{figMBQC} (b) and (c)).
By changing measurement angles, we can generate any state
$U|+\rangle^{\otimes N}$ on the last layer of the cluster state 
(Fig.~\ref{figMBQC} (d)),
where $U$ is any $N$-qubit unitary operator.
(Actually, what we create on the last layer is 
\begin{eqnarray*}
\Big(\bigotimes_{j=1}^NX_j^{x_j}Z_j^{z_j}\Big)U|+\rangle^{\otimes N},
\end{eqnarray*}
where $x_j,z_j\in\{0,1\}$ are random binaries.
We call $\bigotimes_{j=1}^NX_j^{x_j}Z_j^{z_j}$
the byproduct, and we say that we can implement
$U|+\rangle^{\otimes N}$
up to a byproduct.
Since $x_j,z_j$ are determined by measurement results,
we can correct byproduct.
)

\section{Measuring Alice protocol}
\label{sec:MA}
In this section, we will review 
MA protocol~\cite{topoveri,MABQC}.
Let $\rho_{in}$ and $U$ be the input and the program (unitary operation)
of Alice's computation, respectively.
In other words, Alice wants to implement $U$ on $\rho_{in}$,
and obtain $U\rho_{in}U^\dagger$.
If Alice's input is classical, i.e., she knows the classical description of 
the input quantum state $\rho_{in}$, or her input is a classical data,
she does not need to start with $\rho_{in}$:
she can start with the standard state, such as $|0\rangle^{\otimes n}$, 
which is prepared by Bob,
and the preparation of the initial state can be included in $U$.

Bellow, we will introduce two MA protocols.
We will first explain a simpler one: 
a protocol without the verification.
We will next explain a protocol with verification. 
These two protocols satisfy the device-independent blindness.
Note that the device-independent verifiability is not guaranteed
in the second protocol,
because, as we will see later, 
the device-independent verifiability is impossible:
a malicious device can always cheat Alice by pretending that
all tests are passed. 

\subsection{Protocol without verification}
\label{subsec:MA1}
\begin{figure}[htbp]
\begin{center}
\includegraphics[width=0.4\textwidth]{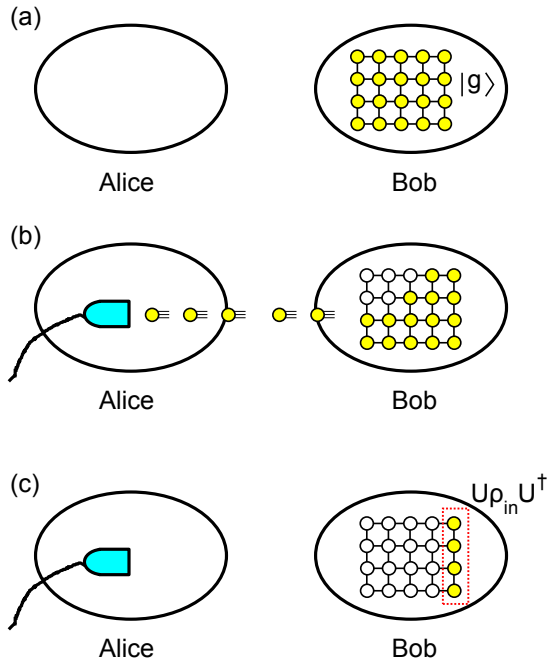}
\end{center}
\caption{
MA protocol without verification.
(a) Bob prepares a resource state. If he is honest, he creates $|g\rangle$.
If he is malicious, he might create completely different state.
(b) He sends each particle to Alice one by one.
Alice measures each particle according to her program.
(c) The last layer of Bob's resource state is $U\rho_{in}U^\dagger$
(up to byproduct)
if he is honest.
} 
\label{MA1}
\end{figure}

Let us consider 
MBQC between two parties,
Alice and Bob (Fig.~\ref{MA1}): Bob first prepares a resource 
state $|g\rangle$, such as the cluster state,
in his laboratory (Fig.~\ref{MA1} (a)).
He next sends each particle to Alice one by one,
and Alice measures each particle in a certain angle which is determined
by her program $U$ (Fig.~\ref{MA1} (b)). The program is kept secret to Bob.

If Bob behaves honestly, i.e., generates the correct resource state
and sends each particle correctly, the last layer
of his resource state becomes
$U\rho_{in}U^\dagger$ (up to byproducts) (Fig.~\ref{MA1} (c)).
If Bob sends it to Alice,
Alice can obtain the correct quantum outcome
(or if she needs the classical outcome,
she can obtain the correct result by measuring it).

Since there is no message transmission from Alice to Bob,
the no-signaling principle guarantees that
Bob cannot learn anything about Alice's input,
measurement angles (i.e., program), and outputs, whatever he does
on his system~\cite{topoveri,MABQC}.
(As we said, we ignore trivial leakage of Alice's information.
In this case, Bob can know the upper bound of the size of Alice's MBQC,
since he creates the resource state.)

This protocol also satisfies the
device-independent blindness, which means that whatever Alice's device does,
Bob cannot learn Alice's secret.
This is again shown by using the no-signaling principle:
due to the no-signaling principle, Alice cannot send any 
message to Bob whatever she does.
Therefore, even if her measuring device does not work correctly,
Bob cannot gain any information about Alice's secret~\cite{topoveri,MABQC,bug}.

\subsection{Protocol with verification}
\label{subsec:MA2}
The above protocol does not satisfy the verifiability.
In other words, if Bob behaves maliciously, Alice accepts a wrong result
although Bob cannot learn Alice's secret.
In order to achieve the verifiability, we modify the above
protocol in the following manner.

Let us define the $N$-qubit state
\begin{eqnarray*}
|\Psi_P\rangle\equiv P\Big(|g\rangle\otimes|+\rangle^{\otimes N/3}\otimes|0\rangle^{\otimes N/3}\Big),
\end{eqnarray*}
where $|g\rangle$ is an $\frac{N}{3}$-qubit universal resource
state for MBQC,
$|+\rangle\equiv\frac{1}{\sqrt{2}}(|0\rangle+|1\rangle)$,
and $P$ is an $N$-qubit permutation, which keeps
the order of qubits in $|g\rangle$. 
(Since Alice does not have any quantum memory,
the order of particles in $|g\rangle$ should not be permutated.)
This permutation is randomly chosen by Alice and kept secret to Bob.

\begin{figure}[htbp]
\begin{center}
\includegraphics[width=0.4\textwidth]{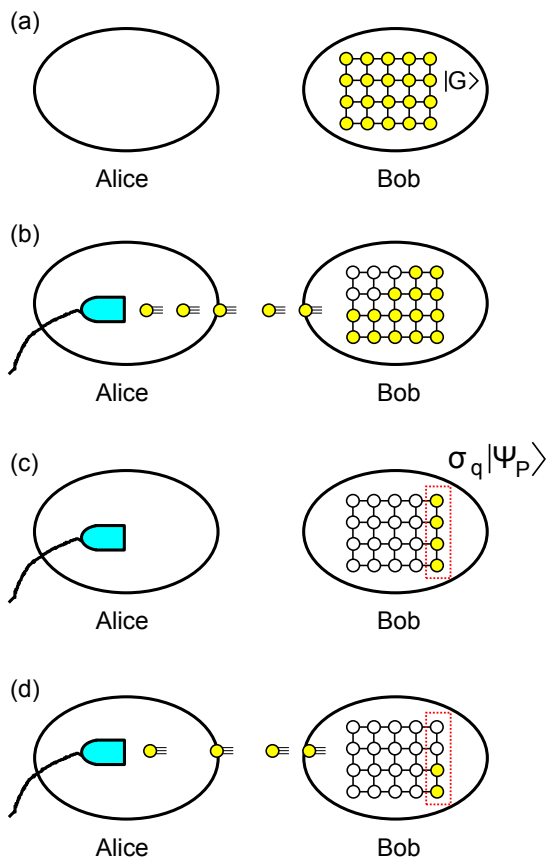}
\end{center}
\caption{
MA protocol with verification.
(a) Bob prepares a resource state. If he is honest, he creates $|G\rangle$.
If he is malicious, he might create a completely different state.
(b) He sends each particle to Alice one by one.
Alice measures each particle to create $\sigma_q|\Psi_P\rangle$.
(c) The last layer of Bob's resource state is $\sigma_q|\Psi_P\rangle$
if he is honest.
(d) Bob sends each particle to Alice one by one.
Alice measures each particle to perform computation 
and to test traps.
} 
\label{MA2}
\end{figure}

%Alice also randomly chooses a parameter
%$q\equiv(x_1,...,x_N,z_1,...,z_N)\in\{0,1\}^{2N}$,
%and keeps it secret to Bob.
Bob prepares a sufficiently large universal
resource state $|G\rangle$ in his laboratory (Fig.~\ref{MA2} (a)).
He sends each qubit of it to Alice one by one,
and Alice measures each qubit (Fig.~\ref{MA2} (b)) until she creates
the $N$-qubit state $\sigma_q|\Psi_P\rangle$
in Bob's laboratory,
where
\begin{eqnarray*}
\sigma_q\equiv \bigotimes_{j=1}^NX_j^{x_j}Z_j^{z_j}
\end{eqnarray*}
with $q\equiv(x_1,...,x_N,z_1,...,z_N)\in\{0,1\}^{2N}$
is the byproduct of MBQC~\cite{Raussendorf} (Fig.~\ref{MA2} (c)).
Here, $X_j$ and $Z_j$ are Pauli operators acting on $j$th qubit.
Throughout this paper, we assume that there is no communication
channel from Alice to Bob.
Then, due to the no-signaling principle, Bob cannot learn anything
about $q$ and $P$~\cite{MABQC}. 
If Bob can learn something about $P$,
Alice can transmit some message to Bob by encoding her message
into $P$, which contradicts to
the no-signaling principle. 
Furthermore, if Bob can learn something about $q$,
Alice can exploit this fact to transmit her message to Bob.

Bob sends each qubit of $\sigma_q|\Psi_P\rangle$ to Alice
one by one,
and Alice does MBQC
on $\sigma_q|\Psi_P\rangle$ with correcting $\sigma_q$ (Fig.~\ref{MA2} (d)).
This means that before measuring $j$th qubit of $\sigma_q|\Psi_P\rangle$
she applies $\sigma_q^\dagger|_j$ on $j$th qubit,
where
$\sigma_q^\dagger|_j$ is the restriction
of $\sigma_q^\dagger$ on $j$th qubit.
For example, $(I\otimes XZ\otimes Z)|_2=XZ$.
Qubits belonging to $|g\rangle$ are used for 
MBQC to realize the unitary
$U$ on the input $\rho_{in}$.
Note that this computation is done with a quantum error correcting
code with the code distance $d$.
States $|0\rangle$ and $|+\rangle$
are used as ``traps"~\cite{FK}.
In other words,
she measures $Z$ on $|0\rangle$ and $X$ on $|+\rangle$,
and if she obtains a minus result,
she rejects the result of the computation.
If results are plus for all traps, she accepts the result
of the computation.

If malicious Bob wants to deviate from the above protocol,
he might apply some operations on his system, or
even he generates completely different state $G'$
in stead of $|G\rangle$.
In this case, he changes traps with high probability,
since he does not know the place of traps.
In particular, if Bob wants to change the logical qubit
of Alice's computation, he has to access at least $d$ qubits,
which increases the probability of changing traps.
It was shown in Ref.~\cite{topoveri}
that
the probability that the logical state is changed and
no trap is flipped
is at most $\left(\frac{2}{3}\right)^{\frac{d}{3}}$,
where $d$ is the distance of the quantum error correcting code.
In other words, the probability that Alice accepts
a wrong result is exponentially small in $d$.
(By doing the concatenation, $d$ can be any large integer.
In particular, it can be sufficiently large for a given security
parameter $\epsilon$, which is introduced later.)
In this way, the verifiability is achieved in MA
protocol.

This protocol also satisfies the device-independent
blindness due to the no-signaling principle.
However, note that the device-independent verifiability
is not satisfied, because a malicious device can always cheat Alice
by outputting plus
results for all trap tests.

%%%%%%%%%%%%%%%%%%%%%%%%%%%%%%%%%%%%%%%%%
\section{Composable security}
\label{sec:comp}
In this section, we will explain the basics of the
composable security~\cite{MR11,Mau11,Unr09,Unr04,Canetti,BPW,UCQKD,BM02,BM04}
concentrating on our setup, namely two-party protocols
with one always-honest client, Alice, and 
one possibly-malicious server, Bob.
Although we will use the constructive 
cryptography~\cite{MR11,Mau11},
similar results may be obtained in the framework of Ref.~\cite{Unr09,Unr04}.

In the framework of the constructive cryptography~\cite{MR11,Mau11},
a protocol $\pi$ is represented by an engine
which has input and output ports,
and performs some functionality.
Protocols implement (approximate) the ideal functionality $S$
by using a resource $R$.
The ideal functionality $S$ has a switch, $f$:
$f=0$ corresponds to honest Bob and $f=1$ corresponds
to malicious Bob.

\begin{figure}[htbp]
\begin{center}
\includegraphics[width=0.45\textwidth]{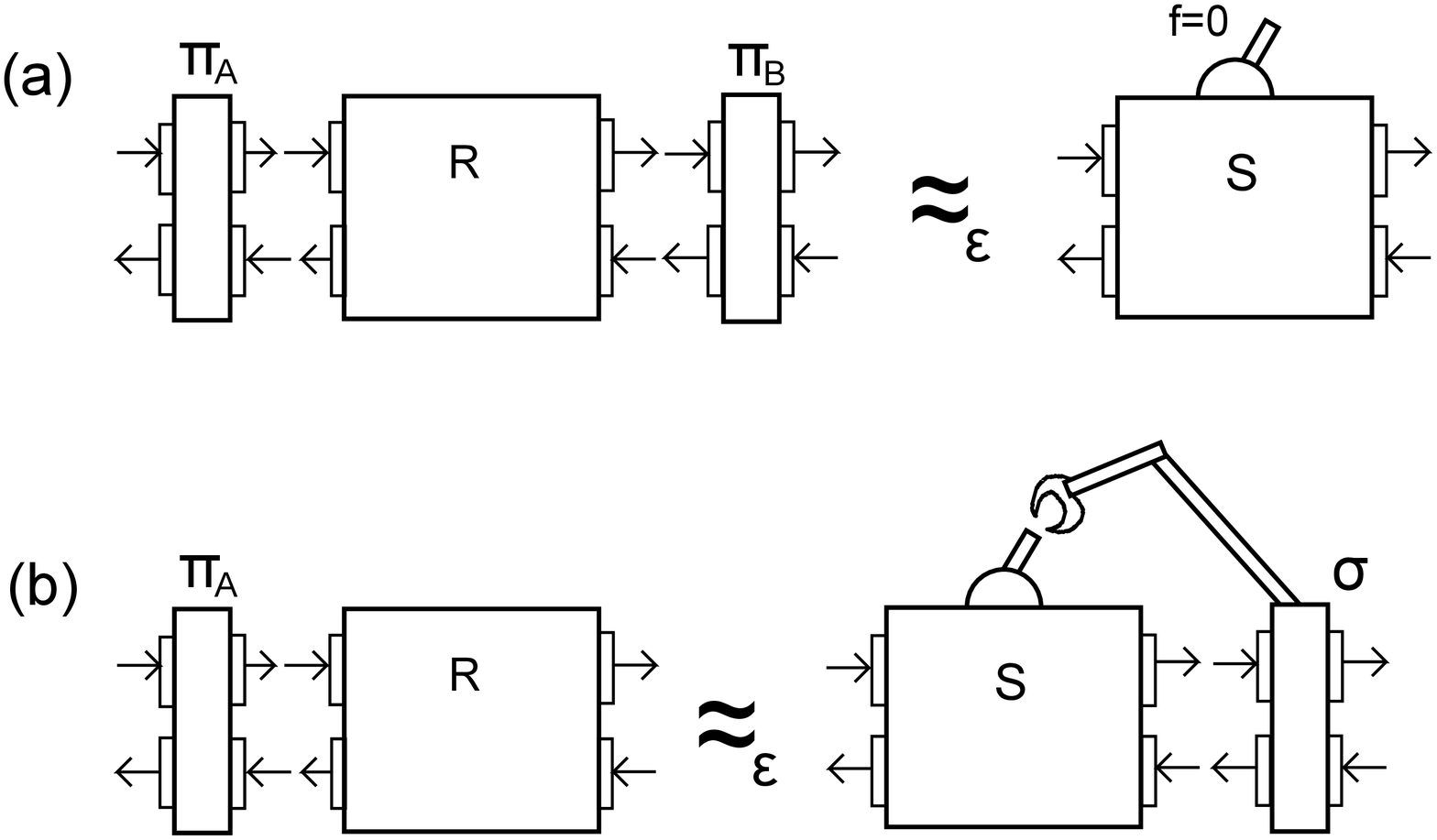}
\end{center}
\caption{
(a) A resource $R$ and protocols $\pi_A$ and $\pi_B$.
(b) An ideal functionality $S$ and a simulator $\sigma$. 
}
\label{AC}
\end{figure}

For example, as is shown in Fig.~\ref{AC} (a), 
Alice's protocol $\pi_A$ and Bob's protocol $\pi_B$
interact (exchange inputs and outputs) with the resource $R$.
If the combination $\pi_AR\pi_B$, which is considered
as a new resource, is $\epsilon$-close
to $S_{f=0}$, 
\begin{eqnarray}
\pi_AR\pi_B\approx_{\epsilon} S_{f=0},
\label{correct}
\end{eqnarray}
we say
that $\pi_A$, $\pi_B$, and $R$
are $\epsilon$-composable correct.
(We will see later why we say ``composable".)
Here, $\epsilon$-close is defined by the diamond-norm as 
\begin{eqnarray*}
\max_{\rho}\|(I\otimes \pi_AR\pi_B)\rho-(I\otimes S_{f=0})\rho\|_{tr}
\le \epsilon,
\end{eqnarray*}
where $\|O\|_{tr}$ is the trace norm of an operator $O$,
and we assume that all inputs and outputs of $\pi_AR\pi_B$ and $S$
are quantum states. (Classical information is encoded in
orthogonal quantum states.)

On the other hand, as is shown in Fig.~\ref{AC} (b),
if there exists an engine $\sigma$, which we call a simulator,
such that $\pi_AR$ is $\epsilon$-close to $S\sigma$,
\begin{eqnarray}
\pi_AR\approx_\epsilon S\sigma,
\label{goal}
\end{eqnarray}
we say that $\pi_A$ and $R$ are $\epsilon$-composable secure.
(We will see later why we say ``composable".)
Again, the $\epsilon$-close means
\begin{eqnarray*}
\max_{\rho}\|(I\otimes \pi_AR)\rho-(I\otimes S\sigma)\rho\|_{tr}
\le\epsilon,
\end{eqnarray*}
where we assume that all inputs and outputs of $\pi_AR$ and $S\sigma$
are quantum states.

Equation~(\ref{goal}) has two very important meanings.
Firstly, it gives a clear definition of security: it is a closeness
of the real protocol to the ideal functionality.
Secondly, as we will explain later, it also guarantees the 
secure composition of the protocol.

\begin{figure}[htbp]
\begin{center}
\includegraphics[width=0.4\textwidth]{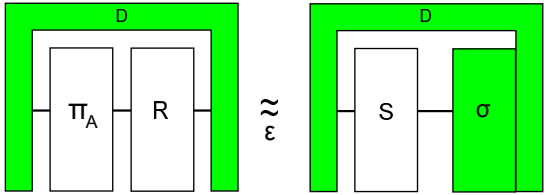}
\end{center}
\caption{
The illustration of Eq.~(\ref{goal}).
} 
\label{imi}
\end{figure}

Let us first explain why Eq.~(\ref{goal}) defines the security.
As is shown in Fig.~\ref{imi},
Eq.~(\ref{goal}) suggests that
the environment (distinguisher),
which interacts with $\pi_A R$ or $S\sigma$,
cannot distinguish $\pi_A R$ and $S\sigma$ within $\epsilon$.
Equation~(\ref{goal}) also suggests that
for any attack $D$ by the distinguisher 
against $\pi_A R$, there exists an attack $D\sigma$ against $S$ 
which causes the same effects to the distinguisher within $\epsilon$.
Therefore,
if $\pi_A R$ is not secure against an attack $D$,
$S$ is insecure against the attack $D\sigma$,
since the distinguisher cannot distinguish $\pi_A R$ and $S\sigma$
(what the distinguisher gains are the same).
However, it contradicts to the assumption that $S$ is secure
against any attack.
Therefore, $\pi_AR$ is secure against any attack within $\epsilon$.

Equation~(\ref{goal})
also means another important concept, the composable security.
If we denote Eq.~(\ref{goal}) by 
\begin{eqnarray*}
R\xrightarrow{\pi_A,\epsilon}S,
\end{eqnarray*}
we can show
(for a proof, see Appendix)
\begin{eqnarray}
R\xrightarrow{\pi,\epsilon}S~\mbox{and}~S\xrightarrow{\pi',\epsilon'}T
&\Rightarrow&
R\xrightarrow{\pi'\circ\pi,\epsilon+\epsilon'}T,\label{comp1}\\
R\xrightarrow{\pi,\epsilon}S~\mbox{and}~R'\xrightarrow{\pi',\epsilon'}S'
&\Rightarrow&
R\| R'\xrightarrow{\pi|\pi',\epsilon+\epsilon'}S\|S'\label{comp2}.
\end{eqnarray}
These equations mean the composability of the protocol.
A protocol might be secure if we use it only a single time in an
isolated environment. Such a security is called the stand-alone
security. However, if the protocol is used in a subroutine
of a larger protocol, the security of the entire protocol
is no longer guaranteed. The above two equations
guarantee the security in such a composable setting.

\begin{figure}[htbp]
\begin{center}
\includegraphics[width=0.3\textwidth]{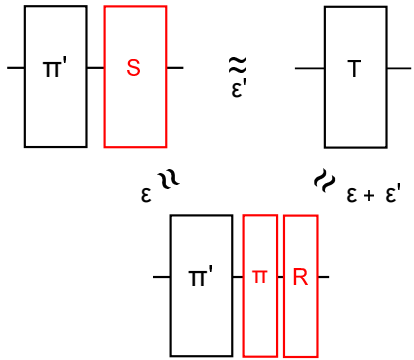}
\end{center}
\caption{
The illustration of Eq.~(\ref{comp1}).
} 
\label{composable}
\end{figure}

Equation~(\ref{comp1}) means the following (Fig.~\ref{composable}).
Let us assume that
we can realize an ideal functionality $T$ by using
a protocol $\pi'$ and a resource $S$ up to the error $\epsilon'$. 
Furthermore, we also assume that
the resource $S$ can be realized by using a protocol $\pi$
and a resource $R$ up to the error $\epsilon$. 
Then, we can realize $T$ by using the composition $\pi'\circ \pi$ 
and $R$ up to the error $\epsilon+\epsilon'$.

Equation~(\ref{comp2}) means that if we run a protocol 
$\pi$ up to the error $\epsilon$
and a protocol $\pi'$ up to the error $\epsilon'$ in parallel,
we can realize the ideal functionality $S|S'$
up to the error $\epsilon+\epsilon'$.

Now it is clear why we call Eq.~(\ref{goal}) the ``composable" security.
The composable correctness is also understood in a similar way.

%%%%%%%%%%%%%%%%%%%%%%%%%%%%%%%%%%%%%%%%%%%%
\section{Composable security of the MA protocol without verification}
\label{sec:result1}
In this section, we will show our first result,
the composable security of the MA protocol
without verification, which was explained in Sec.~\ref{subsec:MA1}.
%Let us define a function $Q$, 
%\begin{eqnarray*}
%\rho_{out}=Q(\rho_{in},U,g',w),
%\end{eqnarray*}
%where $\rho_{out}$, $\rho_{in}$, $g'$ are quantum states,
%$U$ is a unitary operator, and $w$ is an element of the set $W$
%of the descriptions of behaviors of Alice's device.
%$W$ contains 0, and $w=0$ corresponds to the case where
%Alice's device works correctly.
%Therefore, we require that if $w=0$,
%$\rho_{out}$ is the outcome
%of MBQC with input $\rho_{in}$, program $U$, and resource state $g'$.

\begin{figure}[htbp]
\begin{center}
\includegraphics[width=0.35\textwidth]{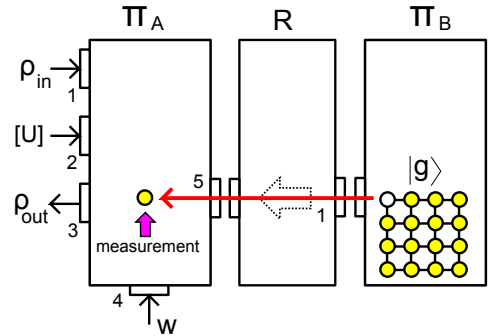}
\end{center}
\caption{
Alice's protocol $\pi_A$, Bob's protocol $\pi_B$,
and the resource $R$ (one-way quantum channel).}
\label{piR2}
\end{figure}

Alice's protocol $\pi_A$,
Bob's protocol $\pi_B$, and the resource $R$ (a one-way quantum channel)
are illustrated in Fig.~\ref{piR2}.
$\pi_A$ accepts 
\begin{itemize}
\item
the input $\rho_{in}$ from the first port,
\item
the classical description $[U]$ of the program $U$
from the second port,
\item
the description $w$ of the ``behavior of Alice's 
device" from the fourth port, 
\item
a state $g'$ from the fifth port
($g'=|g\rangle\langle g|$ if Bob is honest).
\end{itemize}
$\pi_B$ generates the resource state $|g\rangle$ of the measurement-based
quantum computation, and sends each qubit to $\pi_A$
through $R$.

$\pi_A$ runs as follows. 
\begin{itemize}
\item[1.]
If $w=0$, Alice's device works correctly. 
In other words, $\pi_A$ performs MBQC
with input $\rho_{in}$, program $[U]$, and resource state $g'$.
$\pi_A$ then outputs the outcome $\rho_{out}$ of the MBQC
from the third port.
\item[2.]
If $w\neq0$, Alice's device does some wrong behavior specified by $w$.
In this case, $\pi_A$ generates $\rho_{out}$ according to
$\rho_{in}$, $[U]$, $g'$, and $w$,
and outputs it from the third port.
\end{itemize}

\begin{figure}[htbp]
\begin{center}
\includegraphics[width=0.4\textwidth]{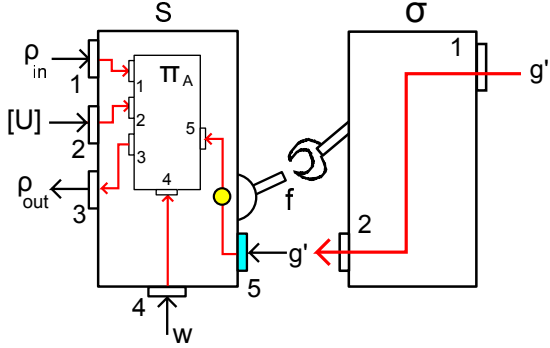}
\end{center}
\caption{
The ideal functionality $S$
and the simulator $\sigma$.
$S$ simulates $\pi_A$ in its inside.
The yellow circle means that the element is not directly
forwarded: if $f=0$, $|g\rangle$ is input to the fifth port
of the simulated $\pi_A$, whereas if $f=1$ $g'$ is forwarded
to the fifth port of the simulated $\pi_A$.
}
\label{Ssigma2}
\end{figure}

The ideal functionality $S$ and the simulator $\sigma$ 
are illustrated in Fig.~\ref{Ssigma2}.
The filtered port (fifth port) is colored
in blue.
$\sigma$ sets $f=1$, accepts a state $g'$ from the first port,
and outputs it from the second port.

$S$ simulates $\pi_A$ in its inside.
$S$ runs as follows.
\begin{itemize}
\item
$S$ accepts the input $\rho_{in}$ from its first port,
and forwards it to the first port of the simulated $\pi_A$.
\item
$S$ accepts the classical description $[U]$ of the program $U$
from its second port,
and forwards it to the second port of the simulated $\pi_A$.
\item
$S$ accepts $w$ from its fourth port,
and forwards it to the fourth port of the simulated $\pi_A$.
\item
If $f=0$, $S$ inputs $|g\rangle$ into the
fifth port of the simulated $\pi_A$.
If $f=1$, $S$ accepts a state $g'$ from the fifth port,
and forwards it to the fifth port of the simulated $\pi_A$.
\item
$S$ gets $\rho_{out}$ from the third port of
the simulated $\pi_A$, and outputs it
from $S$'s third port.
\end{itemize}

Let us first check the correctness.
If $f=0$, $S$ inputs $|g\rangle$ into the
fifth port of the simulated $\pi_A$. 
It is equivalent
to the simulation of $\pi_AR\pi_B$,
since what $R\pi_B$ does is also inputting
$|g\rangle$ into the fifth port of $\pi_A$.
Therefore,
\begin{eqnarray*}
\pi_{A}R\pi_B=S_{f=0}.
\end{eqnarray*}
Hence we obtain the $\epsilon$-composable correctness
with $\epsilon=0$.

Next let us show the composable device-independent blindness.
What $S\sigma$ does is inputting a state $g'$ into the
fifth port of $\pi_A$, which is equivalent to the work of $\pi_AR$.
Therefore,
\begin{eqnarray*}
\pi_AR=S\sigma.
\end{eqnarray*}
Hence we obtain the $\epsilon$-composable device-independent blindness
with $\epsilon=0$.

%%%%%%%%%%%%%%%%%%%%%%%%%%%%%%%%%%%%%%%%
\section{Composable security of MA protocol with verification}
\label{sec:result2}
In this section, we show our second result,
the composable security of the MA protocol with verification,
which was explained in Sec.~\ref{subsec:MA2}.
%Let us define a function $Q'$, 
%\begin{eqnarray*}
%(\rho_{out},e)=Q'(\rho_{in},U,G',w),
%\end{eqnarray*}
%where $\rho_{out}$, $\rho_{in}$, $G'$ are quantum states,
%$U$ is a unitary operator, $e\in\{0,1\}$ is the flag bit
%which indicates whether all trap tests are passed or not,
%and $w$ is an element of the set $W$
%of the descriptions of behaviors of Alice's device.
%$W$ contains 0, and $w=0$ corresponds to the case where
%Alice's device works correctly.

\begin{figure}[htbp]
\begin{center}
\includegraphics[width=0.35\textwidth]{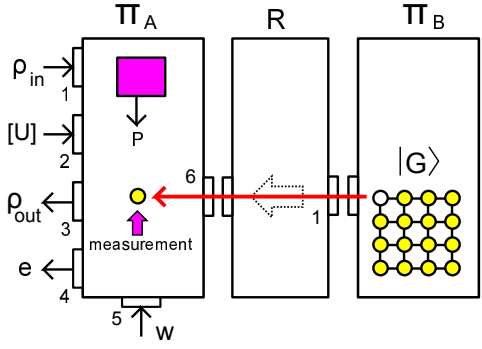}
\end{center}
\caption{
Alice's protocol $\pi_A$, Bob's protocol $\pi_B$, and the resource $R$
(a one-way quantum channel).}
\label{piR}
\end{figure}

Alice's protocol $\pi_A$, Bob's protocol $\pi_B$, and the resource $R$
(a one-way quantum channel)
are illustrated in Fig.~\ref{piR}.
$\pi_A$ accepts
\begin{itemize}
\item
the input $\rho_{in}$ of the computation
from the first port,
\item
the classical description $[U]$ of
the program from the second port,
\item
the description $w$ of the behavior
of Alice's device 
from the fifth port,
\item
a state $G'$ from the sixth port
($G'=|G\rangle\langle G|$ if Bob is honest.)
\end{itemize}

Bob's protocol $\pi_B$ generates the resource state $|G\rangle$
and sends each particle
to Alice one by one through $R$.

If $w=0$, Alice's device works correctly. In other words,
$\pi_A$ runs as follows.
\begin{itemize}
\item[1.]
$\pi_A$ generates a random $N$-qubit permutation $P$.
\item[2.]
$\pi_A$ performs the MBQC ${\mathcal M}_P$ on $G'$, 
where ${\mathcal M}_P$ is the MBQC such that
$\sigma_q|\Psi_P\rangle$ is generated if $G'=|G\rangle\langle G|$.
\item[4.]
$\pi_A$ performs the computation $U$ on $\rho_{in}$
by using $|g\rangle$.
If Bob is malicious, $|g\rangle$ might be
different state, and then $U$ is not correctly implemented.
$\pi_A$ also checks all trap qubits.
\item[5.]
If all traps pass the test, $\pi_A$ outputs $e=0$ from the fourth port.
If at least one trap does not pass the test, $\pi_A$ outputs $e=1$ from the fourth port.
\item[6.]
$\pi_A$ outputs the output $\rho_{out}$ 
of the computation from the third port.
If Bob is honest, $\rho_{out}=U\rho_{in}U^\dagger$.
If he is malicious $\rho_{out}$ might be different state.
\end{itemize}

If $w\ne0$, Alice's device does some wrong behavior
specified by $w$. Then, $\pi_A$ runs as follows.
\begin{itemize}
\item[1.]
$\pi_A$ generates 
$\rho_{out}$ and $e$ according to
$\rho_{in}$, $[U]$, $G'$, and $w$. 
\item[2.]
$\pi_A$ outputs $\rho_{out}$ from the third port, and $e$ from the fourth port.
\end{itemize}

\begin{figure}[htbp]
\begin{center}
\includegraphics[width=0.45\textwidth]{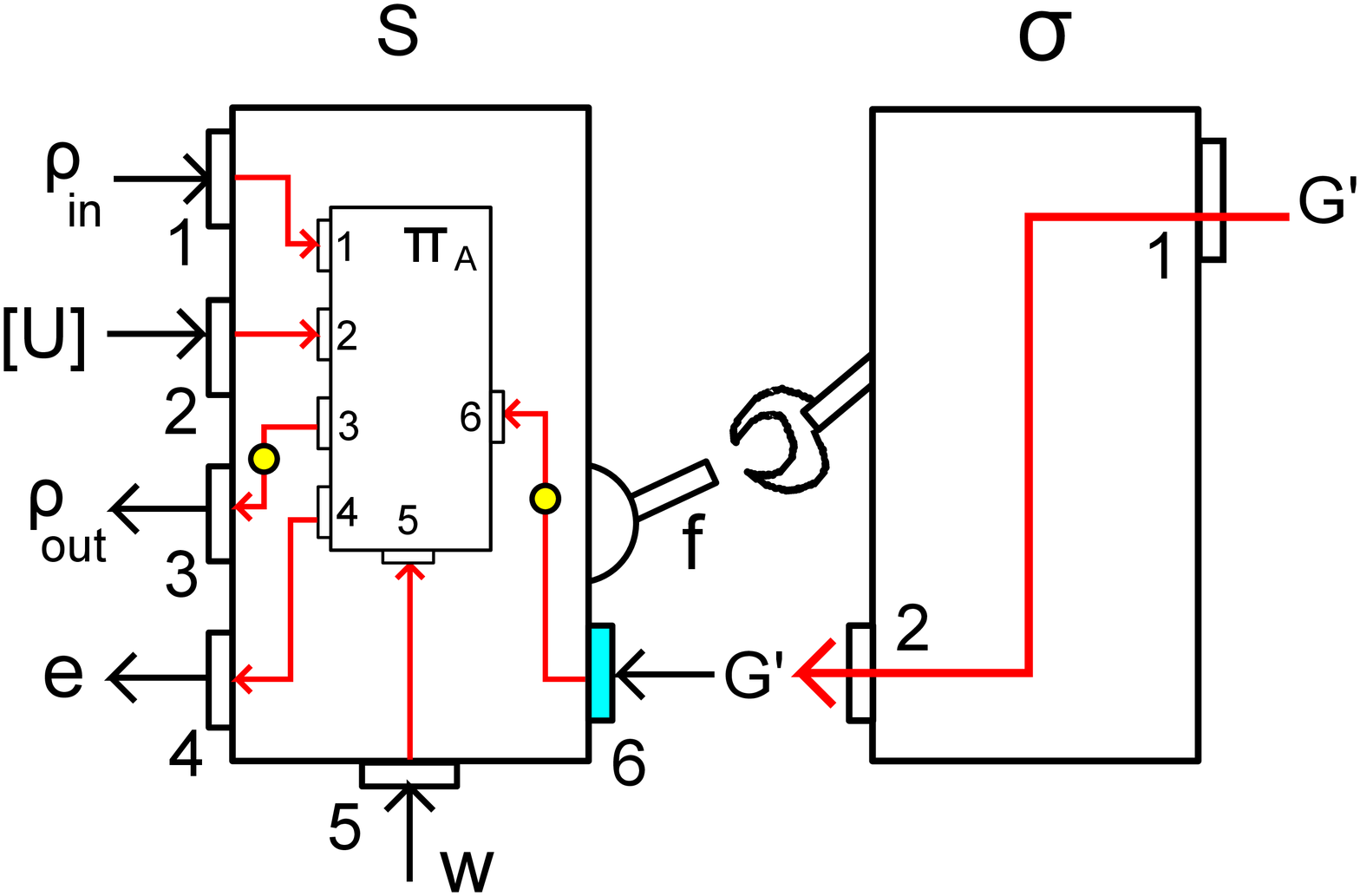}
\end{center}
\caption{
The ideal functionality $S$
and the simulator $\sigma$.
Yellow circles means that elements are not directly transfered:
there are filters.}
\label{Ssigma}
\end{figure}

The ideal functionality $S$ and the simulator $\sigma$
are illustrated in Fig.~\ref{Ssigma}.
The filtered port is colored in blue.
$\sigma$ accepts a state $G'$ from the first port
and outputs it from the second port.
$\sigma$ switches $f=1$.

$S$ simulates $\pi_A$ in its inside.
$S$ runs as follows. 
\begin{itemize}
\item
$S$ accepts the input $\rho_{in}$ from its first port,
and forwards it to the first port of the simulated $\pi_A$.
\item
$S$ accepts the program $[U]$ from its second port,
and forwards it to the second port of the simulated $\pi_A$.
\item
$S$ accepts the description $w$ from its fifth port,
and forwards it to the fifth port of the simulated $\pi_A$.
\item
If $f=0$, $S$ inputs $|G\rangle$
into the sixth port of the simulated $\pi_A$.
If $f=1$, $S$ accepts a state $G'$ from its sixth port, and 
forwards it to the sixth port of the simulated $\pi_A$.
\item
$S$ gets $e$ from the fourth port of the simulated $\pi_A$,
and outputs it from $S$'s fourth port.
\item
$S$ gets $\rho_{out}$ from the third port of the simulated $\pi_A$.
\item
If $w\neq0$, $S$ outputs $\rho_{out}$ from $S$'s third port.
If $w=0$, $S$ works as follows:
\begin{itemize}
\item 
If $e=0$, $S$ outputs $U\rho_{in}U^\dagger$
from $S$'s third port.
\item
If $e=1$, $S$ outputs $\rho_{out}$
from $S$'s third port.
\end{itemize}
\end{itemize}

Let us first show the correctness, Eq.~(\ref{correct}). 
We first consider the case $w=0$.
$\pi_AR\pi_B$ always outputs $e=0$ and $U\rho_{in}U^\dagger$.
On the other hand,
$S_{f=0}$ inputs $|G\rangle$ into the sixth port of the simulated $\pi_A$.
Then the simulated $\pi_A$ always outputs
$e=0$ and $U\rho_{in}U^\dagger$.
Because $e=0$, $S$ always outputs $U\rho_{in}U^\dagger$ from its third port.
Therefore,
$\pi_AR\pi_B=S_{f=0}$.

We next consider the case $w\neq0$.
In this case, 
the output $\rho_{out}$ from the third port of the simulated $\pi_A$ is 
directly output from the third port of $S$.
If $f=0$, $S$ inputs $|G\rangle$ into the sixth 
port of the simulated $\pi_A$, which is equivalent to
the work of $R\pi_B$.
Therefore, again we have shown 
$\pi_AR\pi_B=S_{f=0}$.
In short, we have shown
the $\epsilon$-composable correctness with $\epsilon=0$.

Now let's move on to the security.
Our goal is to show 
Eq.~(\ref{goal}).
%In this case, we say the protocol $\pi_A$ provides 
%$\epsilon$-blind-verifiability. 
In the following, we will show
that Eq.~(\ref{goal}) is satisfied for $\epsilon=2\delta$,
where $\delta$
is the exponentially small probability
that Alice accepts a wrong outcome in the verifiable MA protocol.

\subsection{$w=0$}
\begin{figure}[htbp]
\begin{center}
\includegraphics[width=0.45\textwidth]{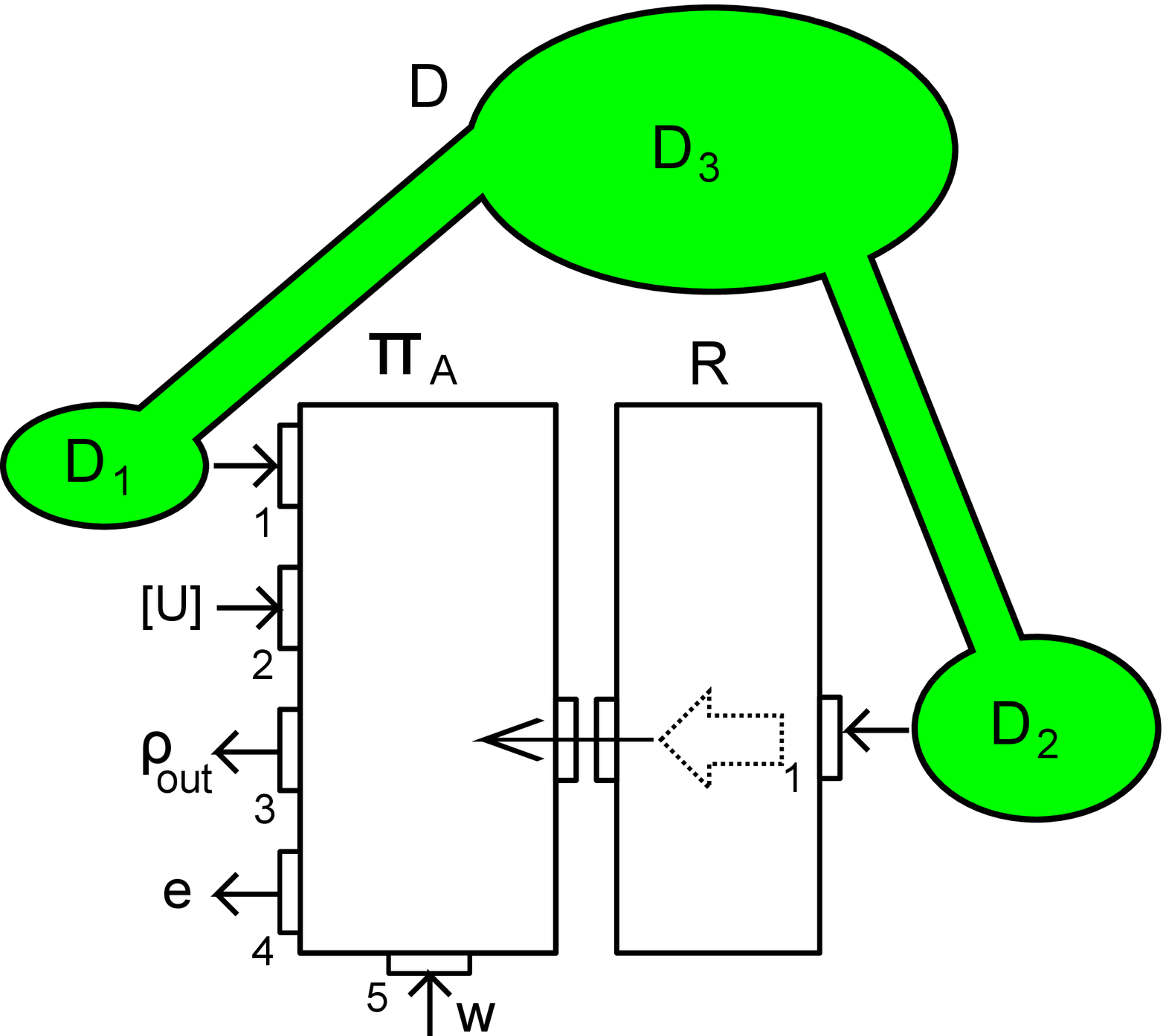}
\end{center}
\caption{
How the distinguisher attacks against $\pi_AR$.
} 
\label{distinguisher1}
\end{figure}

First let us assume $w=0$.
As is shown in Fig.~\ref{distinguisher1}, 
the distinguisher prepares a system $D$,
and divides it into three subsystems, $D=D_1+D_2+D_3$.
The distinguisher inputs $D_1$ into the first port of $\pi_A$,
and $D_2$ into the first port of $R$.
The distinguisher also inputs $[U]$ into the second port of $\pi_A$.
Let $e_0$ and $e_1$ are two orthogonal states which represents
$e=0$ and $e=1$, respectively.
If $D_2=|G\rangle$, $\pi_AR$ outputs
\begin{eqnarray*}
\Big[(U_{D_1}\otimes I_{D_3})\mbox{Tr}_{D_2}(D)(U_{D_1}^\dagger\otimes I_{D_3})
\Big]
\otimes e_0,
\end{eqnarray*}
where $U_{D_1}$ means $U$ acts on $D_1$,
$I_{D_3}$ means $I$ acts on $D_3$,
and
$\mbox{Tr}_{D_2}(D)$ means the partial trace of $D$ over $D_2$.

If $D_2\neq |G\rangle$, $\pi_AR$ outputs
\begin{eqnarray*}
&&\alpha \eta
\otimes e_1
+
\delta\eta_{error} \otimes e_0\\
&&+
(1-\alpha-\delta)
\Big[(U_{D_1}\otimes I_{D_3})\mbox{Tr}_{D_2}(D)(U_{D_1}^\dagger\otimes I_{D_3})
\Big]
\otimes e_0,
\end{eqnarray*}
where $\eta$ is a certain state which
distinguisher gets if $e=1$, $\eta_{error}$
is the state which distinguisher gets when
more than $d$ qubits of the resource state
are affected by errors during the computation,
$0<\delta<1$ is an exponentially small number,
and
$0\le\alpha<1$.

\begin{figure}[htbp]
\begin{center}
\includegraphics[width=0.45\textwidth]{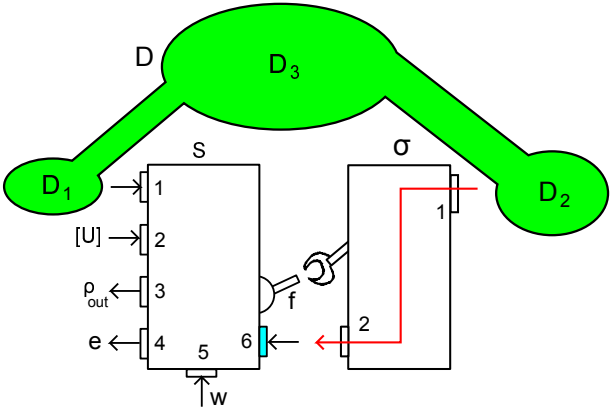}
\end{center}
\caption{
How the distinguisher attacks against $S\sigma$.
} 
\label{distinguisher2}
\end{figure}

On the other hand, as is shown in Fig.~\ref{distinguisher2}, let us assume that
the distinguisher inputs
$D_1$ into the first port of $S$, $D_2$
into the first port of $\sigma$, and
$[U]$ into the second port of $S$.
If $D_2=|G\rangle$, 
$S\sigma$ outputs
\begin{eqnarray*}
\Big[(U_{D_1}\otimes I_{D_3})\mbox{Tr}_{D_2}(D)(U_{D_1}^\dagger\otimes I_{D_3})
\Big]
\otimes e_0.
\end{eqnarray*}

If $D_2\neq|G\rangle$, 
$S\sigma$ outputs
\begin{eqnarray*}
&&\alpha\eta
\otimes e_1\\
&&+(1-\alpha)
\Big[(U_{D_1}\otimes I_{D_3})\mbox{Tr}_{D_2}(D)(U_{D_1}^\dagger\otimes I_{D_3})
\Big]
\otimes e_0.
\end{eqnarray*}

Therefore, the distance between $\pi_AR$ and $S\sigma$
is upper bounded by
\begin{eqnarray*}
\Big\|
\delta\eta_{error}-
\delta
(U_{D_1}\otimes I_{D_3})\mbox{Tr}_{D_2}(D)(U_{D_1}^\dagger\otimes I_{D_3})
\Big\|_{tr}
\le 2\delta,
\end{eqnarray*}
which shows Eq.~(\ref{goal}) with $\epsilon=2\delta$.

\subsection{$w\neq0$}
Next let us consider the case, $w\neq0$.
In this case, 
the output $\rho_{out}$ of the third port of the simulated $\pi_A$
is directly output from the third port of $S$.
Therefore, $S\sigma$ works in the same way as
$\pi_AR$:
$\pi_AR=S\sigma$.

%%%%%%%%%%%%%%%%%%%%%%%%%%%%%%%%%
\section{Discussion}
\label{sec:discussion}
In this paper, we have shown the composable security
of the measuring Alice protocol
by using the constructive cryptography.

In Ref.~\cite{composable},
authors introduced the definition of 
the stand-alone $\epsilon$-blind verifiability,
and showed it is equivalent to the composable $\epsilon$-blind verifiability
(Theorem 5.3 of Ref.~\cite{composable}).
They also defined the stand-alone $\epsilon$-blindness,
the stand-alone $\epsilon$-verifiability,
and the stand-alone $\bar{\epsilon}$-independent
$\epsilon$-verifiability.
They showed that the stand-alone blindness
and the independent verifiability means 
the composable security (Lemma 6.6, Theorem 6.7, and Corollary 6.8).
Since the FK protocol~\cite{FK} satisfies these individual stand-alone
definitions, they conclude that FK protocol is composable secure (Appendix C
of Ref.~\cite{composable}).
It might be possible to show the composable blind-verifiability
of MA protocol by showing in a similar way,
i.e., first showing that MA protocol satisfies
the above individual stand-alone definitions
and then use Theorem 6.7 of Ref.~\cite{composable}.
However, these individual definitions are introduced for
the BFK-type setup, i.e., Alice generates some states and exchanges
quantum states and classical messages between Bob.
Therefore directly showing the composable
security of MA protocol,
which we have done in this paper,
seems to be easier and more transparent.
It would be a subject of future work 
to investigate the relation between MA protocol
and the above individual stand-alone definitions.

%Authors of Ref.~\cite{composable},
%show the composable blindness of BFK protocol by directly
%showing the existence of a simulator $\sigma$, which satisfies
%$\pi_A R \pi_B\approx_{\epsilon}S\sigma$.

\acknowledgements
TM is supported by the program to disseminate
tenure tracking system by MEXT.
TK is supported by JSPS KAKENHI Grant Numbers
23236071, 24240001, 23650004, and 24106008.

\appendix*
\section{}
For simplicity, we omit the identity operator $I$.
\begin{eqnarray*}
&&\max_{\rho}\|(\pi' \pi R)\rho-(T\sigma'\sigma)\rho\|_{tr}\\
&=&
\|(\pi' \pi R)\eta-(T\sigma'\sigma)\eta\|_{tr}\\
&=&
\|(\pi' \pi R)\eta-(\pi'S\sigma)\eta+(\pi' S\sigma)\eta-
(T\sigma'\sigma)\eta\|_{tr}\\
&\le&
\|(\pi' \pi R)\eta-(\pi'S\sigma)\eta\|_{tr}
+\|(\pi' S\sigma)\eta-(T\sigma'\sigma)\eta\|_{tr}\\
&=&
\|{\mathcal E}[(\pi R)\eta^*]-{\mathcal E}[(S\sigma)\eta^*]\|_{tr}\\
&&+\|{\mathcal F}[(\pi' S)\eta^{**}]-{\mathcal F}[(T\sigma')\eta^{**}]\|_{tr}\\
&\le&
\epsilon+\epsilon',
\end{eqnarray*}
where $\eta$, $\eta^*$, and $\eta^{**}$ are certain states,
and ${\mathcal E}$ and ${\mathcal F}$ are certain CPTP maps.
If we consider $\sigma'\sigma$ as the simulator for $T$, this shows
Eq.~(\ref{comp1}).

\begin{eqnarray*}
&&\max_{\rho}\|(\pi R\otimes \pi' R')\rho
-(S\sigma\otimes S'\sigma')\rho
\|_{tr}\\
&=&
\|(\pi R\otimes \pi' R')\eta
-(S\sigma\otimes S'\sigma')\eta
\|_{tr}\\
&=&
\|(\pi R\otimes \pi' R')\eta
-(\pi R\otimes S'\sigma')\eta\\
&&+(\pi R\otimes S'\sigma')\eta
-(S\sigma\otimes S'\sigma')\eta
\|_{tr}\\
&\le&
\|(\pi R\otimes \pi' R')\eta
-(\pi R\otimes S'\sigma')\eta\|_{tr}\\
&&+\|(\pi R\otimes S'\sigma')\eta
-(S\sigma\otimes S'\sigma')\eta
\|_{tr}\\
&\le&\epsilon'+\epsilon,
\end{eqnarray*}
where $\eta$ is a certain state.
This shows Eq.~(\ref{comp2}).

%%%%%%%%%%%%%%%%%%%%%%%%%%%%%%%%%%%%%%%%


\begin{thebibliography}{99}
\bibitem{bug}
We assume that there is no unwanted leakage of information
from Alice's laboratory to Bob. For example,
Alice's laboratory is not bugged by Bob.
This is a standard assumption in many cryptographic protocols,
such as the quantum key distribution (QKD).
Without such a natural assumption, 
all cryptographic protocols are trivially broken.
\end{thebibliography}
\end{document}